\documentclass[aps,prl,twocolumn,showpacs,amsmath,floatfix]{revtex4}

\usepackage{bm}
\usepackage{epsfig}
\usepackage{psfig}

\newcommand{\nn}{n_1}
\newcommand{\vs}{v_s}

\begin{document}

\title{Collisions of solitons and vortex rings in cylindrical
Bose-Einstein condensates}
\author{Stavros Komineas$^1$ and Joachim Brand$^2$}
\affiliation{$^1$
TCM Group,  Cavendish Laboratory,
Madingley Road, Cambridge CB3 0HE, United Kingdom. \\
$^2$Max Planck Institute for the Physics of Complex Systems,
N\"othnitzer Stra{\ss}e 38, 01187 Dresden, Germany}

\date{\today}

\begin{abstract}
Interactions of solitary waves in a cylindrically confined
Bose-Einstein condensate are investigated by simulating their head-on
collisions.
Slow vortex rings and fast solitons are found to collide elastically
contrary to the situation in the three-dimensional homogeneous Bose gas.
Strongly inelastic collisions are absent for low density condensates
but occur at higher densities for intermediate velocities.
The scattering behaviour is rationalised by use of dispersion diagrams.
During inelastic collisions, spherical shell-like structures
of low density are formed and
they eventually decay into
depletion droplets with solitary wave features.
The relation to similar shells observed in a recent experiment
[Ginsberg {\it et al.} 
Phys Rev.\ Lett.\ {\bf 94}, 040403 (2005)] is discussed.
\end{abstract}

\maketitle


The observation of dark solitons and vortex rings in a series of
experiments in Bose-Einstein condensates (BECs)
\cite{burger,denschlag,dutton,anderson} has shown that this is an
excellent physical system for the study of nonlinear waves.
Theoretical work has shown that the latter have the form of
solitons, vortex rings, and solitonic vortices
\cite{feder,brand1,komineas2,komineas3}.
In the recent
experiment of Ginsberg {\it et al.}~\cite{ginsberg} collisions between
solitary waves were observed.  While vortex rings and solitons were
robust in many collision events, in some cases shell structures of low
particle density were observed, which had not been predicted before.

In the one-dimensional (1D) nonlinear Schr\"odinger equation (NLS),
which describes the 1D Bose gas, soliton collisions are elastic, that
is no energy is radiated and the outgoing solitons are the same as the
colliding ones \cite{ablowitz}.  In the homogeneous three-dimensional
(3D) Bose gas, solitary waves have the form of vortex rings or
rarefaction pulses \cite{jones}.  Their interactions are important
for the understanding of superfluid turbulence. In contrast to the 1D case,
collisions are generally inelastic. Large vortex rings annihilate when they
collide head-on by increasing their radius and radiating phonons
\cite{koplik,berloffJPA}.
Collisions at oblique angles or with an impact 
parameter result in vortex-line reconnections or produce Kelvin-wave
radiation \cite{leadbeater,berloff}.  Solitary wave collisions in
trapped 2D systems were previously considered in
Refs.~\cite{carr,schulte}.

A model system that connects the 1D
and 3D solitary waves is given by the cylindrically confined BEC.
Studying elementary collision processes in this model is thus of
fundamental theoretical interest and furthermore it leads to
a deeper understanding of the experimentally observed structures.

\begin{figure}[ht]
\epsfig{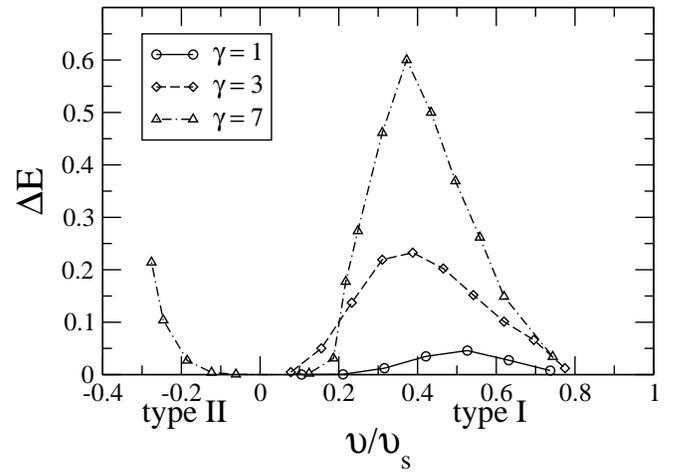}
 \caption{The energy fraction $\Delta E$, defined in (\ref{eq:loss}),
 which is radiated during collision of solitons and vortex rings 
 of type I and II (opposite direction)
 as a function of their initial velocity, for three values of
 the coupling constant $\gamma = n_1 a$.
 The velocity is normalised to the speed of sound $\vs=0.95, 1.29,
 1.61$
(in units of $\hbar/(m l_\rho)$)
 for $\gamma = 1, 3, 7$, respectively.
The symbols correspond to the simulations performed.} 
  \label{fig:loss}
\end{figure}

In this Letter we present a detailed numerical and theoretical
study of head-on collisions between solitary waves in
a cylindrically confined BEC.
We test the robustness of solitary waves when they interact.
The detailed behaviour of solitary waves under collision
for low density condensates is found to be similar to soliton dynamics in the 
NLS \cite{ablowitz}.
This is, however, only a limiting case for the present system, which also shows
more complicated dynamical behaviour.
For higher particle densities we find that the collision
dynamics of solitary waves, including vortex rings,
is very different than in the homogeneous 3D Bose gas.
As shown in Fig.~\ref{fig:loss}, we find elastic collisions for small
and large velocities.
At intermediate energies, however, inelastic collisions
often produce temporary spherical shells
reminiscent of those observed experimentally \cite{ginsberg}.
In the following, we introduce the model and discuss the various
regimes in detail.

We assume a cylindrical and infinitely-elongated trap as
in Ref.~\cite{komineas2}
with symmetry axis z.
The Gross-Pitaevskii equation can be written 
in the dimensionless form
\begin{equation}  \label{eq:gp}
i\, \frac{\partial\Psi}{\partial t} = -\frac{1}{2}\, \Delta\Psi
+\frac{1}{2}\,\rho^2 \,\Psi + 4\pi\gamma\,|\Psi|^2 \Psi,
\end{equation}
where $\rho=\sqrt{x^2+y^2}$ is the radial coordinate. The
dimensionless coupling constant
$\gamma\equiv\nn a$ is the
only parameter entering the equation, where $a$ is the scattering length
and $\nn$ is the linear particle density along the symmetry axis z.
Length is measured in units of the oscillator length
$l_\rho = \sqrt{\hbar/m \omega_\rho}$, where $m$ is the atomic mass
and $\omega_\rho$ is the transverse trapping frequency.
The unit of time is $1/\omega_\rho$.
At $z\to \pm \infty$
the wave function  approaches the ground state in the transverse plane
with  $\partial\Psi/\partial z=0$ and
$\int{|\Psi|^2 d^2x} = 1$.

\begin{figure}
\epsfig{file=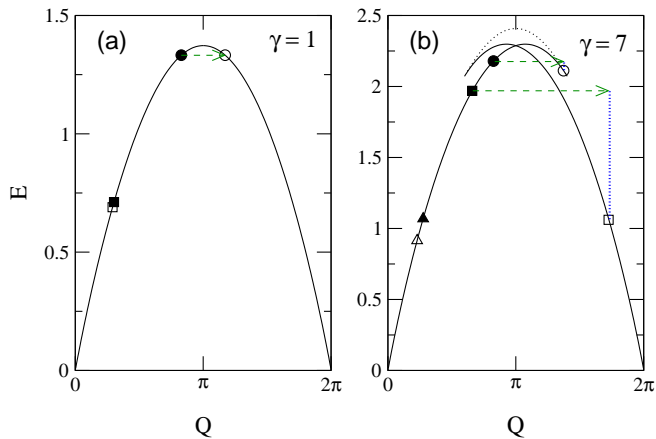,width=8.6cm}
 \caption{The excitation energy $E$ in units of $\nn  \hbar^2 /(m l_\rho)$
 versus impulse $Q$ in units of $\nn \hbar$
for axially symmetric solitary waves \cite{komineas2}.
At $\gamma=1$ in panel (a) there is one family of solitons.
Panel (b)  for $\gamma=7$ shows two fundamental branches (solid lines)
of vortex rings related by (\ref{eq:symmetry}).
The vortex ring radius has a minimum at the cusps, where $|v|= 0.3 \vs$. 
The impulse transfer during bounce-off collisions is indicated by the dashed
arrow connecting initial solitary waves (filled symbols, see text) with
collision products (open symbols).
The vertical dotted lines
 indicate the energy loss.
}
 \label{fig:dispersion}
\end{figure}

The wave functions of solitary waves
in the cylindrical trap are of the form
\begin{equation}
 \Psi(x,y,z,t;v) = \psi(x,y,z-v t)\, e^{-i\mu t},
\end{equation}
where $v$ is the velocity of propagation and $\mu$ the chemical potential.
For $\gamma \lesssim 1.5$ only one family of solitary waves exists with
velocities $-\vs < v < \vs$. They are related to the dark soliton of the NLS.
We show the energy-momentum dispersion for $\gamma=1$
in Fig.~\ref{fig:dispersion}a \cite{komineas2}.
A modified momentum $Q$, called the {\it impulse},
has been defined so that
the velocity is given by the slope of the curve $v=dE/dQ$ \cite{komineas2}.
Solitons with opposite velocities are obtained by the symmetry transformation
\begin{equation}  \label{eq:symmetry}
\Psi \to \Psi^*,\; v \to -v,
\end{equation}
where the star denotes complex conjugation.

We numerically simulate the elementary process of a head-on collision of
two solitons at $\gamma=1$ with wave functions related by
Eq.~(\ref{eq:symmetry}). 
We assume axial symmetry throughout the collision process and we
typically use a lattice size 
$6\times 60$ in the $(\rho,z)$ plane
and a lattice spacing 0.1 for both variables.
We integrate Eq.~(\ref{eq:gp}) in time using a Runge-Kutta method
for various values of velocity $v$ of the initial solitons
each calculated as in Ref.~\cite{komineas3}.
The outgoing solitons after collision are in all cases very similar
to the incoming ones and no visible radiation is produced,
thus collisions are elastic.
A measure of the energy transformed to radiation
during collision is
\begin{equation}  \label{eq:loss}
\Delta E = \frac{E_i - E_f}{E_i},
\end{equation}
where $E_i$ is the energy of the incoming solitons and
$E_f$ is the energy of the outgoing ones.
The energy $E_f$ of the outgoing solitons is inferred by
measuring their velocity.
As seen in Fig.~\ref{fig:loss}, $\Delta E$
reaches a maximum 5\% for
intermediate velocities $v\approx 0.5\vs\, (\vs=0.95)$.

\begin{figure}  
\epsfig{file=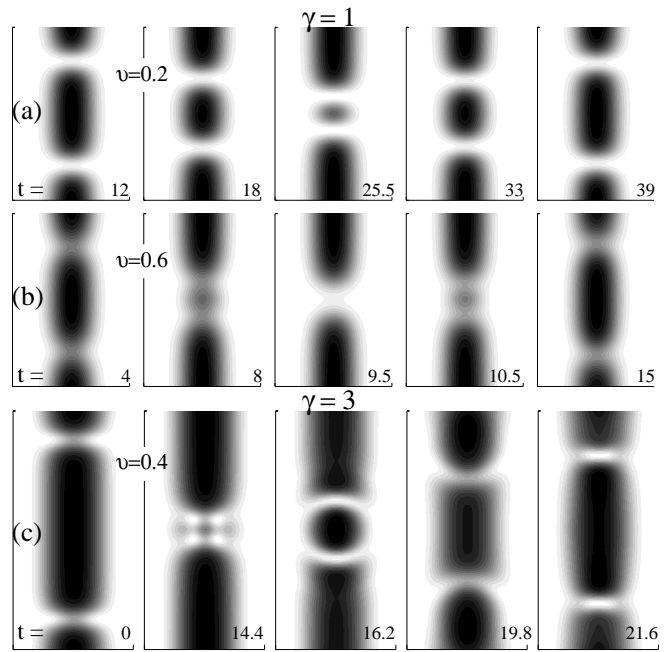,width=8.6cm}
\caption{
Plots of the particle density  $|\Psi|^2$ 
for solitary wave collisions on a scale from white at zero to black at
maximum density for each plot.
Each row shows five snapshots of a solitary wave collision with
values of $\gamma=1$ (a,b), and $\gamma=3$ (c), and
velocities $v$ as indicated.
The time for each snapshot is shown in the figure. 
The spatial dimensions of the plots are $8\times 12$ for (a,b)
and $8\times 16$ for (c).
The initial separation (at time $t=0$) of the
solitary waves is $d=12$ in units of $l_\rho$ for all cases.
}
  \label{fig:density1-3}
\end{figure}

In Fig.~\ref{fig:density1-3}a we present five snapshots
of the simulation for solitons with initial velocity $v=0.2 (0.21 \vs)$.
The solitons decelerate as they approach and interact, they reach a
minimum separation and finally bounce back.
This behaviour is typical for small velocities.
One of the initial solitons is denoted by a filled circle
in Fig.~\ref{fig:dispersion}a,
and it appears that it effectively moves along the dispersion
curve due to the collision process to the point denoted by an open circle
at almost the same energy level.
This picture
gives a precise result for the final outcome of the collision,
and it also describes gross features of the collision process.
It cannot give the detailed features of the process
because the dispersion pertains
only to isolated solitary waves.
As the velocity of the initial solitons
increases, they pass through each other during collision
as is seen in Fig.~\ref{fig:density1-3}b for $v=0.6 (0.63 \vs)$.
In the dispersion diagram of Fig.~\ref{fig:dispersion}a, 
one of the solitons is represented by a filled
square before and by an open square after collision.
The energy difference between the two points is small, thus
the collision is almost elastic. 
The overall picture at low
$\gamma$ closely resembles the elastic soliton collisions in the
integrable NLS \cite{ablowitz}.

We repeat the simulations for a denser condensate
by setting $\gamma=3$ in Eq.~(\ref{eq:gp}).
Vortex rings, albeit with a very inconspicuous ring structure, form now part
of the solitary wave family.
The energy radiated during collision
reaches now a maximum 23\% at $v=0.4\vs$
($\vs=1.29$) as shown in Fig.~\ref{fig:loss}.
Nevertheless, collisions for slow and for fast solitary waves are also in
this case elastic and their behaviour resembles that for
$\gamma=1$ \cite{energy}.
In Fig.~\ref{fig:density1-3}c we show five snapshots of the simulation
for $v=0.4 (0.31\vs)$.
The second snapshot shows the formation of fully fledged
vortex rings at the time of collision which do not exist
as isolated solitary waves.
In fact we have noticed the transient formation
of vorticity during solitary-wave collisions in many cases in this
work. Similar observations were also made in the simulations of the
Ginsberg experiment \cite{ginsberg} and for simulated collisions of
rarefaction pulses with vortex lines in the homogeneous BEC \cite{berloff}.
The outgoing waves are seen to be of the soliton type in the
fourth snapshot. However, these are later spontaneously transformed to
depletion droplets, i.e.\ strongly localised regions of low density,
as seen in the last snapshot.
At later times not shown in the figure, 
the droplets transform into soliton-like structures  and
periodically revive 
two to three times until we can no longer follow the
simulation due to lattice limitations.
The speed of the wave is almost constant at $0.48\vs$.

We further study how the picture changes as the coupling increases.
The energy-momentum dispersion for axially symmetric solitary waves
changes substantially for $\gamma \geq 4$,
and it is shown for $\gamma=7$ in Fig.~\ref{fig:dispersion}b \cite{komineas3}.
It has two fundamental branches (solid lines in the figure)
which are related by Eq.~(\ref{eq:symmetry}),
and they both contain axially symmetric vortex rings
with velocities of both signs.
For example, for the branch originating on the left
we have
$1.61 = \vs \geq v \geq -v_1 =-0.30\vs$ where $-v_1$ is the
velocity at the 
cusp.
For velocities close to $\vs$, the solitary waves have the structure
of gray solitons. For lower velocities they
are vortex rings whose radius decreases
as $v$ decreases (as $v \to -v_1$).
In the following we need
to distinguish between the vortex rings
on the left and those on the right of the dispersion maximum.
We call the vortex rings
with $v > 0$ type I, and those with
$-v_1 \leq v < 0$ type II.
Finally, the symmetry relation (\ref{eq:symmetry}) gives
a second branch in Fig.~\ref{fig:dispersion}b with mirror imaged type
I and II rings. The branch denoted by a dotted line
in Fig.~\ref{fig:dispersion}b contains
unstable solitary waves of higher energy and will not be discussed further here.

We simulate the collision of two counter-propagating vortex rings of
opposite circulation related by Eq.~(\ref{eq:symmetry}).
We first discretized
the wave function in a 3D lattice in Cartesian coordinates.
After checking in test runs that the axial symmetry
was not broken during collision
we proceeded to extensive simulations
assuming axial symmetry throughout the process.
The energy radiated due to collision for a coupling
constant $\gamma=7$ is shown in Fig.~\ref{fig:loss}.
The maximum is 60\% and occurs for initial velocity $v=0.37\vs, (\vs=1.61)$.
On the other hand, collisions at low and high velocities
appear to be elastic, i.e.,
the outgoing solitary waves have almost the same energy as
the incoming ones.
The three curves in Fig.~\ref{fig:loss} for $\gamma=1,3,\,{\rm and}\,7$
show that collisions are elastic for all couplings
when the colliding solitary waves are slow or when their velocity
is close to the speed of sound \cite{energy}.
For intermediate velocities energy is radiated during collision
and this is higher as the coupling increases.

\begin{figure}  
\epsfig{file=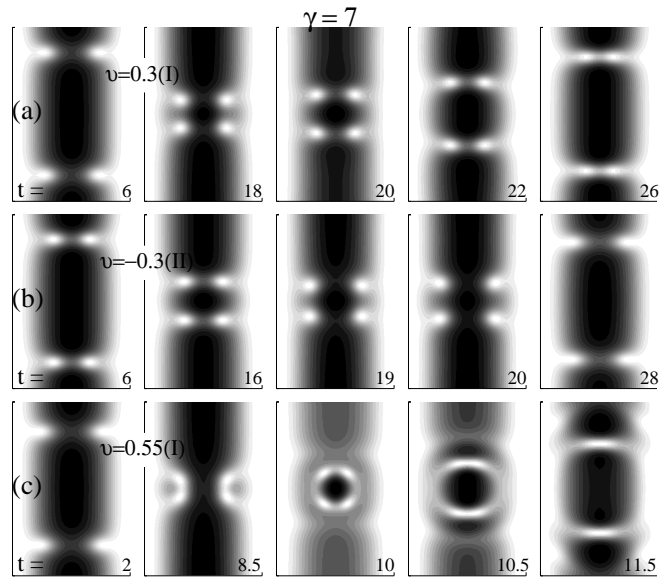,width=8.6cm}
\caption{
Particle density  $|\Psi|^2$ for solitary wave collisions
for $\gamma=7$.
The frames are $8\times 12$.
}
  \label{fig:density7}
\end{figure}

Fig.~\ref{fig:density7}a shows the collision
of vortex rings with an initial velocity $v=0.3 (0.19\vs)$.
The initial rings are denoted by a filled circle in the dispersion
(Fig.~\ref{fig:dispersion}b) and they are type I.
They reach a minimum distance, their radius decreases,
and subsequently they bounce back. The outgoing rings of type II
still have the original circulation
and are denoted by an open circle on  the dispersion curve.
The change in nature of the rings is apparent in the
change of the ring radius between the first and last snapshots
in Fig.~\ref{fig:density7}a.
The energy radiated during this process is 3\%.
It appears that each ring has effectively moved along one of the
branches of the dispersion curve.
This picture gives a precise result for the final outcome of the collision,
but it also gives a faithful representation of the
gross features throughout the collision process.
The almost elastic 
collisions of slow vortex rings seen here are
in stark contrast to collisions in the 3D bulk where vortex rings increase
their radius and eventually annihilate \cite{koplik}. They are
inhibited to do this in our model due to the transverse confinement.

We may choose the initial vortex rings to be type II
but still have initial velocity $|v|=0.3 (0.19\vs)$
(Fig.~\ref{fig:density7}b). In this case the collision
causes them to increase their radius, and they are eventually
transformed to type I rings moving in opposite directions.
The process is essentially the reverse of the one in Fig.~\ref{fig:density7}a.

The arguments pertaining to the dispersion curve can be employed
to rationalise the observed behaviour
for higher velocities (lower energies).
One can thus imagine a collision where the initial type I
vortex rings have energy slightly lower
than the energy at the cusp, as for example at the point 
denoted by a filled square in Fig.~\ref{fig:dispersion}b
(corresponding to $v=0.5 (0.31\vs)$).
The bounce-back process of vortex rings cannot occur in this case.
Instead, intermediate structures are formed, as discussed below,
and the collisions become highly inelastic.
In the example of Fig.~\ref{fig:dispersion}b the final waves,
denoted by an open square, have substantially lower energy than
the initial rings.
This mechanism of inelastic collision suggests that
the maximum energy loss occurs at intermediate velocities.

A further example of a collision for an
initial velocity 0.55 (0.34$\vs)$ for type I rings is shown in
Fig.~\ref{fig:density7}c.
The vortex rings form a shell-like object of low density
as they collide, as seen in the third entry of the
figure.
Since we have assumed axial symmetry, the shell is actually almost spherical,
and strongly reminiscent of those
observed in Ref.~\cite{ginsberg}. However, the shells reported here
are produced
by an elementary head-on collision process which is significantly simpler
than the process reported in the experiment where many
nonlinear wave structures interact simultaneously
while the condensate is expanding.
The resemblance between the observed structures suggests
that inelastic collisions of axisymmetric
nonlinear waves is the fundamental process underlying
the generation of spherical waves observed experimentally \cite{ginsberg}

As the rings move away to opposite directions they form 
depletion droplets (similar to those seen in Fig.~\ref{fig:density1-3}c),
which are shown in the last entry of the figure.
These travel coherently for a distance of approximately 10 units
and they eventually seem to decay into solitons.
The depletion droplets are distinctly different than
solitary waves theoretically studied in a confined BEC but
they are reminiscent of the
rarefaction pulses in the homogeneous Bose gas
\cite{jones,berloffnote}.

The collision in Fig.~\ref{fig:density7}c is highly inelastic
in the sense that the main outgoing
waves carry only part of the total energy while the rest of the energy is
radiated away. However, it is possible that some of the remaining
energy is carried by shallow gray solitons.

At small energies ($v$ close to $\vs$)
the solitary waves have a gray-soliton character. They pass through
each other during collision much like the situation for small couplings.
In Fig.~\ref{fig:dispersion}b the solitary wave with
$v=1.0 (0.62\vs)$ is denoted by a filled triangle. After the collision it
has only slightly lower energy (open triangle).
Thus, collisions for large velocities are elastic as is also shown
in Fig.~\ref{fig:loss} \cite{energy}.

Concluding, we have shown that elastic collisions of
solitary waves can occur in cylindrical BECs.
Shell structures reminiscent of recent
experimental observations were shown to arise already in elementary
inelastic head-on collisions. Possible extensions of the present work
to larger densities, non-axisymmetric solitary waves, and
beyond the head-on case will provide further valuable insights into the
dynamics of nonlinear waves in confined BECs.

We thank L.\ Hau, N.\ Ginsberg, N.\ Berloff and N.\ Papanicolaou for 
inspiring discussions.
This work was supported by EPSRC Grant No GR/R96026/01, and has benefited
from a visit to KITP, Santa Barbara (SK).



\begin{thebibliography}{10}
\bibitem{energy} The numerically determined value of $\Delta E$
is very small for $v\approx \vs$, but it cannot be calculated very accurately
because $E_i$ and $E_f$ are small.
\end{thebibliography}
\end{document}